\documentclass[12pt]{article}

\usepackage[margin=1in,nohead,dvips,a4paper]{geometry}
\usepackage[dvips]{graphics}
\usepackage[small,center]{caption}

\newtheorem{con}{Conjecture}

\def\boe #1{%
\begin{picture}(121,45)
  \put(0,10){\includegraphics{bo6#1.eps}}
  \put(3,5){\makebox(0,0)[c]{\scriptsize 1}}
  \put(26,5){\makebox(0,0)[c]{\scriptsize 2}}
  \put(49,5){\makebox(0,0)[c]{\scriptsize 3}}
  \put(72,5){\makebox(0,0)[c]{\scriptsize 4}}
  \put(95,5){\makebox(0,0)[c]{\scriptsize 5}}
  \put(117,5){\makebox(0,0)[c]{\scriptsize 6}}
\end{picture}%
}

\def\boo #1{%
\begin{picture}(100,42)
  \put(0,10){\includegraphics{bo5#1.eps}}
  \put(3,5){\makebox(0,0)[c]{\scriptsize 1}}
  \put(26,5){\makebox(0,0)[c]{\scriptsize 2}}
  \put(50,5){\makebox(0,0)[c]{\scriptsize 3}}
  \put(74,5){\makebox(0,0)[c]{\scriptsize 4}}
  \put(98,5){\makebox(0,0)[c]{\scriptsize 5}}
\end{picture}%
}

\def\bpe #1{%
\begin{picture}(69,80)
  \put(10,10){\scalebox{.5}{\includegraphics{bp6#1.eps}}}
  \put(37,72){\makebox(0,0)[c]{\scriptsize 1}}
  \put(65,57){\makebox(0,0)[c]{\scriptsize 2}}
  \put(65,20){\makebox(0,0)[c]{\scriptsize 3}}
  \put(37,3){\makebox(0,0)[c]{\scriptsize 4}}
  \put(9,20){\makebox(0,0)[c]{\scriptsize 5}}
  \put(9,57){\makebox(0,0)[c]{\scriptsize 6}}
\end{picture}%
}

\def\bpo #1{%
\begin{picture}(69,80)
  \put(10,10){\scalebox{.5}{\includegraphics{bp7#1.eps}}}
  \put(37,72){\makebox(0,0)[c]{\scriptsize 1}}
  \put(65,57){\makebox(0,0)[c]{\scriptsize 2}}
  \put(69,27){\makebox(0,0)[c]{\scriptsize 3}}
  \put(52,7){\makebox(0,0)[c]{\scriptsize 4}}
  \put(22,7){\makebox(0,0)[c]{\scriptsize 5}}
  \put(6,27){\makebox(0,0)[c]{\scriptsize 6}}
  \put(9,57){\makebox(0,0)[c]{\scriptsize 7}}
\end{picture}%
}

\begin{document}

\title{O(1) loop model with different boundary conditions\\
and symmetry classes of alternating-sign matrices}
\author{A.~V.~Razumov and Yu.~G.~Stroganov\\
\small \it Institute for High Energy Physics\\[-.5em]
\small \it 142280 Protvino, Moscow region, Russia}

\date{}

\maketitle

\begin{abstract}
This work as an extension of our recent paper where we have found a
numerical evidence for the fact that the numbers of the states of the fully
packed loop (FPL) model with fixed link-patterns coincide with the
components of the ground state vector of the dense O$(1)$ loop model for
periodic boundary conditions and an even number of sites. Here we give two
new conjectures related to different boundary conditions. Namely, we
suggest that the numbers of the half-turn symmetric states of the FPL model
with fixed link-patterns coincide with the components of the ground state
vector of the dense O$(1)$ loop model for periodic boundary conditions and
an odd number of sites and that the corresponding numbers of the vertically
symmetric states describe the case of the open boundary conditions and an
even number of sites.
\end{abstract}

\vskip 1em

In paper~\cite{RS} we made some conjectures related to combinatorial
properties of the ground state vector of the XXZ spin chain for the
asymmetry parameter $\Delta=-1/2$ and an odd number of sites. In the
subsequent paper~\cite{BGN} Batchelor, de Gier and Nienhuis considered two
variations of this model along with the corresponding dense O$(n)$ loop
model at $n = 1$ and notably increased the number of models and related
combinatorial objects. Later we made some additional conjectures for the
case of the XXZ spin chain with twisted boundary conditions \cite{RST} and
for the case of the dense O$(1)$ loop model \cite{RSRS}. In the present
paper we continue the consideration of the latter case.

Let us first review the results of papers \cite{BGN, RSRS} and then give
our new conjectures.

The state space of the dense O$(1)$ loop model can be constructed as
follows, see paper~\cite{BN} and references therein. For the open case one
considers $N$ vertices placed on a line. Then for an even $N$ one connects
the vertices pairwise from the same side of the line without intersections,
see table \ref{t1} for $N = 6$.
\begin{table}[ht]
\begin{center}
\begin{tabular}{ccc}
\hline \hline
\multicolumn{2}{c}{\vrule height 1.1em width 0pt} & $\Psi$ \\
\hline
\boe{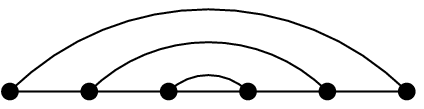} & & \raisebox{10bp}{1} \\ \hline
\boe{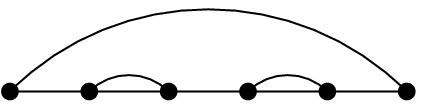} & & \raisebox{10bp}{4} \\ \hline
\boe{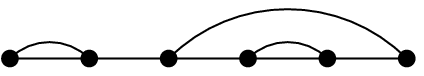} & \boe{4} & \raisebox{10bp}{5} \\ \hline
\boe{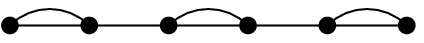} & & \raisebox{10bp}{11} \\
\hline
\end{tabular}
\end{center}
\caption{Basis vectors and components of ground state vector of dense
O$(1)$ loop model with open boundary condition for $N = 6$}
\label{t1}
\end{table}
The state space in question is the vector space of formal linear
combinations with complex coefficients of the obtained pairings.
In the case of an odd $N$ one vertex remains unpaired and this vertex
divides the line into two parts formed by non-itersecting pairings, see
table \ref{t2} for $N = 5$.
\begin{table}[ht]
\begin{center}
\begin{tabular}{cccc}
\hline \hline
\multicolumn{3}{c}{\vrule height 1.1em width 0pt} & $\Psi$ \\
\hline
\boo{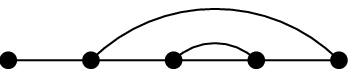} & \boo{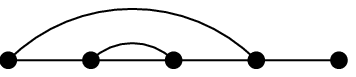} & & \raisebox{10bp}{1} \\ \hline
\boo{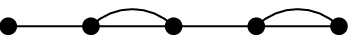} & \boo{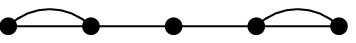} & \boo{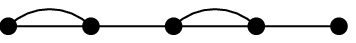} & \raisebox{10bp}{3} \\
\hline
\end{tabular}
\end{center}
\caption{Basis vectors and components of ground state vector of dense
O$(1)$ loop model with open boundary condition for $N = 5$}
\label{t2}
\end{table}
For the periodic case we consider $N$ vertices placed on a circle. Then
for an even $N$ we connect the vertices pairwise inside the circle without
intersections, see table \ref{t3} for $N = 6$.
\begin{table}[ht]
\begin{center}
\begin{tabular}{cccc}
\hline \hline
\multicolumn{3}{c}{\vrule height 1.1em width 0pt} & $\Psi$ \\
\hline
\bpe{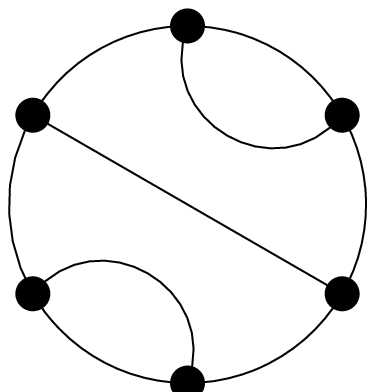} & \bpe{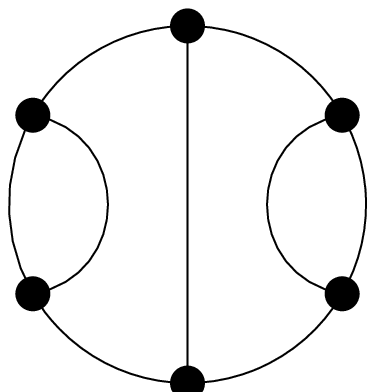} & \bpe{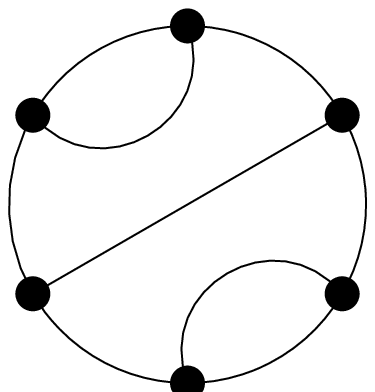} & \raisebox{30bp}{1} \\ \hline
\bpe{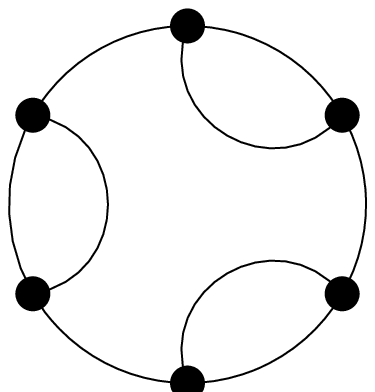} & \bpe{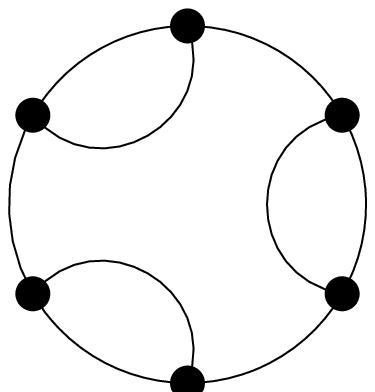} & & \raisebox{30bp}{2} \\
\hline
\end{tabular}
\end{center}
\caption{Basis vectors and components of ground state vector of dense
O$(1)$ loop model with periodic boundary condition for $N = 6$ (up to
rotations)}
\label{t3}
\end{table}
For an odd $N$ one vertex remains again unpaired, see table \ref{t4} for $N
= 7$ where only the parings which do not differ by a rotation are
presented.
\begin{table}[ht]
\begin{center}
\begin{tabular}{ccccc}
\hline \hline
\multicolumn{1}{c}{\vrule height 1.1em width 0pt} & $\Psi$ &
\multicolumn{2}{c}{} & $\Psi$ \\
\hline
\bpo{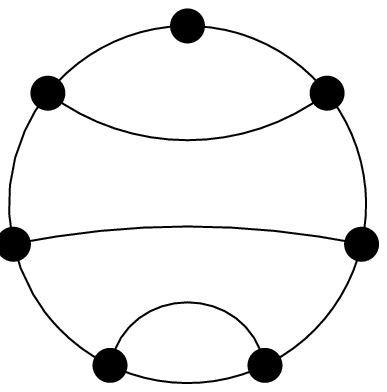}& \raisebox{30bp}{1} & \hspace{1em} \bpo{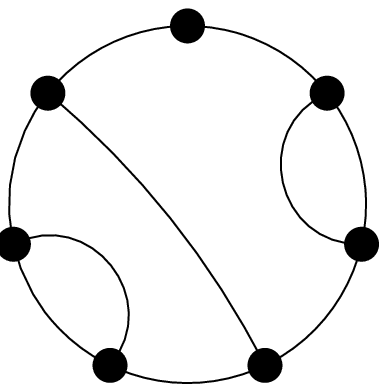}& \bpo{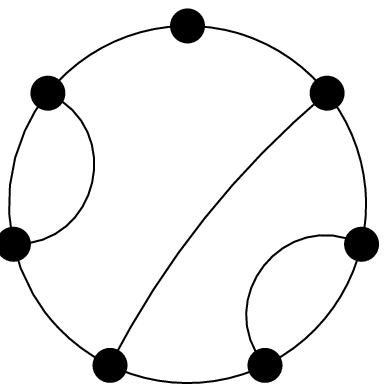}&
\raisebox{30bp}{14} \\ \hline
\bpo{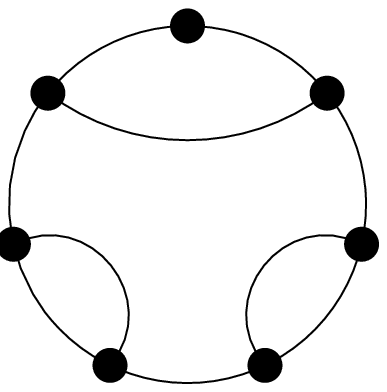}& \raisebox{30bp}{6} & \hspace{1em} \bpo{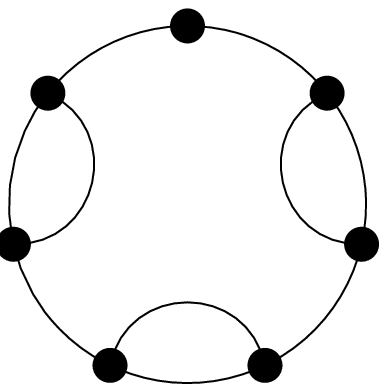}& & \raisebox{30bp}{49}
\\
\hline
\end{tabular}
\end{center}
\caption{Basis vectors and components of ground state vector of dense
O$(1)$ loop model with periodic boundary condition for $N = 7$}
\label{t4}
\end{table}
The state space of the model is again the vector space of formal linear
combination of the parings with complex coefficients.

To construct the Hamiltonian of the dense O$(1)$ loop model let us define
the set of operators $h_i$, where the index $i$ runs from 1 to $N-1$ for
the open case, and from 1 to $N$ for the periodic case.

Consider first the case of an even $N$. For a fixed $i$ take a general
basis vector. Let in this basis vector the $i$-th vertex is paired with
the $(i+1)$-th one. In this case the operator $h_i$ left the basis
vector unchanged. Otherwise, let the $i$-th vertex is paired with the
$j$-th one and the $(i+1)$-th vertex is paired with the $k$-th one. In this
case the operator $h_i$ removes the two pairings under consideration and
pairs the $i$-th vertex with the $(i+1)$-th one and the $j$-th vertex with
the $k$-th one. In the case of an odd $N$ one of the vertices $i$-th and
$(i+1)$-th may be unpaired. Let it be the $i$-th vertex, and let the
$(i+1)$-th vertex is paired with the $k$-th vertex. In this case the
operator $h_i$ removes this pairing and pairs the $i$-th and $(i+1)$-th
vertices. Here the $k$-th vertex becomes unpaired. The case where the
$(i+1)$-th vertex is unpaired is treated similarly.

The Hamiltonian of the dense O$(1)$ loop model is defined as the sum of the
operators $h_i$ taken with the minis sign. Let $H$ be the matrix of the
Hamiltonian. From the definition of the operators $h_i$ it follows that the
sum of the matrix elements belonging to each column of the matrix $H$ is
equal to $-(N-1)$ for the open case, and it is equal to $-N$ for the
periodic case. Therefore, this matrix has a left eigenvector with all
components equal to~$1$. Thus, the Hamiltonian of the system has an
eigenvector with the eigenvalue $-(N-1)$ for the open case, and with the
eigenvalue $-N$ for the periodic case. There is a strong evidence that this
vector is the ground state vector of the model, see, for example,
\cite{RSRS}. The components of the ground state vector for some partial
cases are presented in tables \ref{t1}--\ref{t4}. More examples can be
found in paper \cite{BGN}.

The periodic case with an even number of sites is in a sense the simplest
one. It was conjectured in paper~\cite{BGN} that under some appropriate
normalization the sum of the components of the ground state vector for
$N=2n$ is equal to the number of $n \times n$ alternating sign matrices
(ASMs), usually denoted by $A_n$. We supposed that the values of the ground
state vector components correspond to some subclasses of the
ASMs~\cite{RSRS}. To define this correspondence one has to use the specific
`reincarnation' of the ASMs which we will discuss now.

\begin{figure}[ht]
  \centering
  \begin{minipage}[t]{.425\linewidth}
    \centering
    \begin{picture}(126,126)
      \put(5,5){\scalebox{.8}{\includegraphics{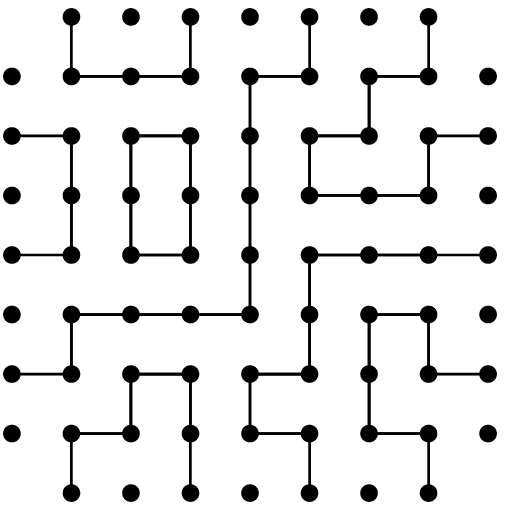}}}
      \put(104,126){\makebox(0,0)[c]{\scriptsize 4}}
      \put(125,91){\makebox(0,0)[c]{\scriptsize 5}}
      \put(125,63){\makebox(0,0)[c]{\scriptsize 6}}
      \put(125,36){\makebox(0,0)[c]{\scriptsize 7}}
      \put(105,1){\makebox(0,0)[c]{\scriptsize 8}}
      \put(77,1){\makebox(0,0)[c]{\scriptsize 9}}
      \put(49,1){\makebox(0,0)[c]{\scriptsize 10}}
      \put(21,1){\makebox(0,0)[c]{\scriptsize 11}}
      \put(0,36){\makebox(0,0)[c]{\scriptsize 12}}
      \put(0,63){\makebox(0,0)[c]{\scriptsize 13}}
      \put(0,91){\makebox(0,0)[c]{\scriptsize 14}}
      \put(21,126){\makebox(0,0)[c]{\scriptsize 1}}
      \put(49,126){\makebox(0,0)[c]{\scriptsize 2}}
      \put(77,126){\makebox(0,0)[c]{\scriptsize 3}}
    \end{picture}
    \caption{One of the possible states of the FPL model for $n=7$}
    \label{gfpl7}
  \end{minipage}
  \hspace{.05\linewidth}
  \begin{minipage}[t]{.425\linewidth}
    \centering
    \begin{picture}(97,97)
      \put(5,5){\scalebox{.8}{\includegraphics{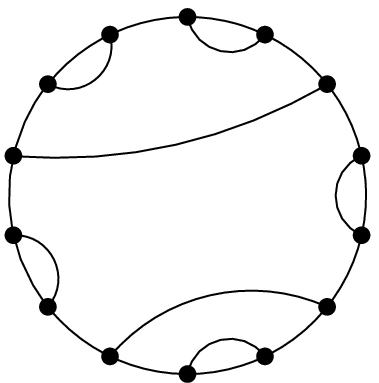}}}
      \put(49,98){\makebox(0,0)[c]{\scriptsize 1}}
      \put(70,94){\makebox(0,0)[c]{\scriptsize 2}}
      \put(87,80){\makebox(0,0)[c]{\scriptsize 3}}
      \put(97,60){\makebox(0,0)[c]{\scriptsize 4}}
      \put(97,38){\makebox(0,0)[c]{\scriptsize 5}}
      \put(87,18){\makebox(0,0)[c]{\scriptsize 6}}
      \put(70,5){\makebox(0,0)[c]{\scriptsize 7}}
      \put(49,0){\makebox(0,0)[c]{\scriptsize 8}}
      \put(28,5){\makebox(0,0)[c]{\scriptsize 9}}
      \put(10,18){\makebox(0,0)[c]{\scriptsize 10}}
      \put(1,38){\makebox(0,0)[c]{\scriptsize 11}}
      \put(0,60){\makebox(0,0)[c]{\scriptsize 12}}
      \put(10,80){\makebox(0,0)[c]{\scriptsize 13}}
      \put(28,94){\makebox(0,0)[c]{\scriptsize 14}}
    \end{picture}
    \caption{The pairing-pattern corresponding to figure \ref{gfpl7}}
    \label{gp7}
  \end{minipage}
\end{figure}

The background information on the ASMs and their different combinatorial
forms can be found in the recent review by Propp \cite{Propp} and in
references therein. The most important for us is the bijection of the ASMs
and the states of the fully packed loop (FPL) model which can be described
as follows. Following paper \cite{Propp} define the `generalized
tic-tac-toe' graph as the graph formed by $n$ horizontal lines and $n$
vertical lines meeting $n^2$ intersections of degree $4$, with $4n$
vertices of degree $1$ at the boundary. Then we number the vertices of
degree $1$. We start with the left top vertex and number clockwise every
other vertex. Now consider subgraphs of the  underlying tic-tac-toe graph
such that each of the $n^2$ internal vertices lies on exactly two of the
selected edges and each numbered external vertex lies on a selected edge,
while each unnumbered external vertex does not lie on a selected edge (see,
for example, figure \ref{gfpl7}).
These subgraphs are the states of the FPL model. They are in bijective
correspondence with the ASMs. In particular, the number of such
states is equal to the number of ASMs. Each state of the FPL model define a
so-called pairing-pattern describing the pairings of the external vertices.
We depict such a pattern as a circle with $2n$ vertices placed on it and
connected pairwise inside the circle without intersection, see figure
\ref{gp7}. Having in mind the relation to ASMs we denote the number of the
states of the FPL model corresponding to the pairing-pattern $\pi$ by
$A_n(\pi)$. Denote the set of all possible pairing-patterns by $\Pi_n$. It
is evident that we can identify any pairing-pattern with the corresponding
basis vector of the dense O$(1)$ loop model. Then the main conjecture of
our paper \cite{RSRS} can be formulated as

\begin{con}
For the case of an even $N = 2n$ and the periodic boundary conditions the
vector
\[
\Psi = \sum_{\pi \in \Pi_n} \pi A_n(\pi)
\]
is the ground state vector of the dense {\rm O(1)} loop model.
\end{con}

Proceed now to the periodic case with an odd $N = 2n + 1$. In paper
\cite{BGN} a formula for the sum of the components of the ground state
vector was conjectured for this case. It appears that this formula
actually gives the number of $(2n + 1) \times (2n + 1)$ half-turn
symmetric ASMs, which we denote by $A^{\mathrm{HT}}_{2n+1}$. Information
about various symmetry classes of alternating sign matrices can be found in
papers by Robbins \cite{R} and Kuperberg \cite{K} and in references
therein.

\begin{figure}[ht]
  \centering
  \begin{minipage}[t]{.425\linewidth}
    \centering
    \begin{picture}(126,126)
      \put(5,5){\scalebox{.8}{\includegraphics{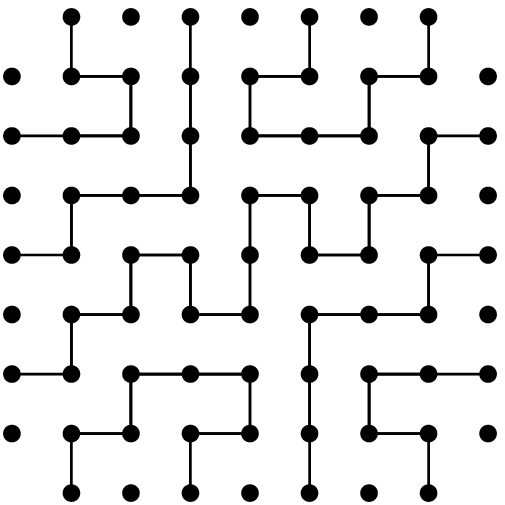}}}
      \put(104,126){\makebox(0,0)[c]{\scriptsize 4}}
      \put(125,91){\makebox(0,0)[c]{\scriptsize 5}}
      \put(125,63){\makebox(0,0)[c]{\scriptsize 6}}
      \put(125,36){\makebox(0,0)[c]{\scriptsize 7}}
      \put(105,1){\makebox(0,0)[c]{\scriptsize $1'$}}
      \put(77,1){\makebox(0,0)[c]{\scriptsize $2'$}}
      \put(49,1){\makebox(0,0)[c]{\scriptsize $3'$}}
      \put(21,1){\makebox(0,0)[c]{\scriptsize $4'$}}
      \put(0,36){\makebox(0,0)[c]{\scriptsize $5'$}}
      \put(0,63){\makebox(0,0)[c]{\scriptsize $6'$}}
      \put(0,91){\makebox(0,0)[c]{\scriptsize $7'$}}
      \put(21,126){\makebox(0,0)[c]{\scriptsize 1}}
      \put(49,126){\makebox(0,0)[c]{\scriptsize 2}}
      \put(77,126){\makebox(0,0)[c]{\scriptsize 3}}
    \end{picture}
    \caption{One of the possible half-turn symmetric states of the FPL
      model for $n=7$}
    \label{gfplht7}
  \end{minipage}
  \hspace{.05\linewidth}
  \begin{minipage}[t]{.425\linewidth}
    \centering
    \begin{picture}(185,97)
      \put(5,5){\scalebox{.8}{\includegraphics{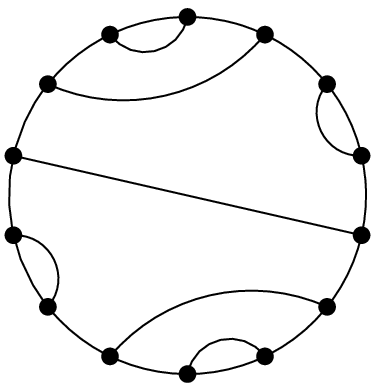}}}
      \put(49,98){\makebox(0,0)[c]{\scriptsize 1}}
      \put(70,94){\makebox(0,0)[c]{\scriptsize 2}}
      \put(87,80){\makebox(0,0)[c]{\scriptsize 3}}
      \put(97,60){\makebox(0,0)[c]{\scriptsize 4}}
      \put(97,38){\makebox(0,0)[c]{\scriptsize 5}}
      \put(87,18){\makebox(0,0)[c]{\scriptsize 6}}
      \put(70,5){\makebox(0,0)[c]{\scriptsize 7}}
      \put(49,0){\makebox(0,0)[c]{\scriptsize $1'$}}
      \put(28,5){\makebox(0,0)[c]{\scriptsize $2'$}}
      \put(10,18){\makebox(0,0)[c]{\scriptsize $3'$}}
      \put(1,38){\makebox(0,0)[c]{\scriptsize $4'$}}
      \put(0,60){\makebox(0,0)[c]{\scriptsize $5'$}}
      \put(10,80){\makebox(0,0)[c]{\scriptsize $6'$}}
      \put(28,94){\makebox(0,0)[c]{\scriptsize $7'$}}
      \put(130,25){\scalebox{.4}{\includegraphics{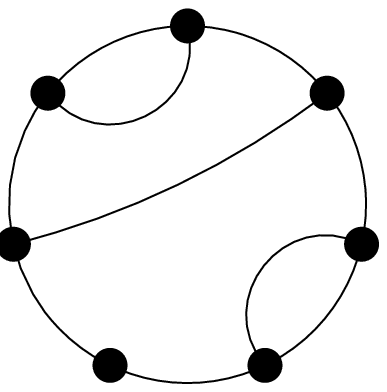}}}
      \put(151,76){\makebox(0,0)[c]{\scriptsize 1}}
      \put(172,67){\makebox(0,0)[c]{\scriptsize 2}}
      \put(179,43){\makebox(0,0)[c]{\scriptsize 3}}
      \put(163,22){\makebox(0,0)[c]{\scriptsize 4}}
      \put(141,22){\makebox(0,0)[c]{\scriptsize 5}}
      \put(123,43){\makebox(0,0)[c]{\scriptsize 6}}
      \put(130,67){\makebox(0,0)[c]{\scriptsize 7}}
    \end{picture}
    \caption{The pairing-pattern and the basis vector corresponding to
      figure \ref{gfplht7}}
    \label{gpht7}
  \end{minipage}
\end{figure}

It can be shown that the bijection of the ASMs and the states of the FPL
model sends a half-turn symmetric ASM into a half-turn symmetric state and
vice versa. Here by a half-turn symmetric state we mean a state whose
picture rotated on $180^\circ$ over its center coincides with itself, see
figure \ref{gfplht7} for an example.
The corresponding pairing-pattern also has the half-turn symmetry. Denote
the set of such pairing-patterns by $\Pi^{\mathrm HT}_{2n+1}$, and
the number of the half-turn symmetric ASMs corresponding to the
pairing-pattern $\pi \in \Pi^{\mathrm{HT}}_{2n+1}$ by
$A^{\mathrm{HT}}_{2n+1}(\pi)$.

It is possible to establish a bijection of the set $\Pi^{\mathrm
HT}_{2n+1}$ and the basis of the dense O$(1)$ loop model for the case under
consideration. To this end we use a special numbering of the vertices,
which should be clear from consideration of figure \ref{gfplht7}. The
bijection in question is realised via identifying $i$-th vertex with the
$i'$-th one, see figure \ref{gpht7}. Identifying the states of the
dense O$(1)$ loop model with the half-turn symmetric pairing-patterns we
formulate our first new conjecture as follows.

\begin{con}
For the case of an odd $N = 2n+1$ and the periodic boundary conditions the
vector
\[
\Psi = \sum_{\pi \in \Pi^{\mathrm{HT}}_{2n+1}} \pi
A^{\mathrm{HT}}_{2n+1}(\pi)
\]
is the ground state vector of the dense {\rm O(1)} loop model.
\end{con}

Now consider the open case for an even $N = 2n$. It was conjectured in
paper \cite{BGN} that in this case the sum of the components of the ground
state vector coincides with the number of $(2n+1) \times (2n+1)$ vertically
symmetric ASMs, which we denote by $A^{\mathrm V}_{2n+1}$.

\begin{figure}[ht]
  \centering
  \begin{minipage}[t]{.425\linewidth}
    \centering
    \begin{picture}(126,126)
      \put(5,5){\scalebox{.8}{\includegraphics{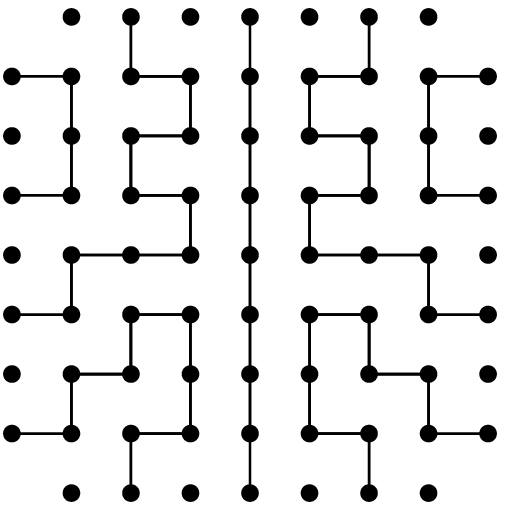}}}
      \put(125,105){\makebox(0,0)[c]{\scriptsize 2}}
      \put(125,77){\makebox(0,0)[c]{\scriptsize 3}}
      \put(125,49){\makebox(0,0)[c]{\scriptsize 4}}
      \put(125,22){\makebox(0,0)[c]{\scriptsize 5}}
      \put(91,0){\makebox(0,0)[c]{\scriptsize 6}}
      \put(63,1){\makebox(0,0)[c]{\scriptsize $0'$}}
      \put(36,1){\makebox(0,0)[c]{\scriptsize $6'$}}
      \put(1,23){\makebox(0,0)[c]{\scriptsize $5'$}}
      \put(1,50){\makebox(0,0)[c]{\scriptsize $4'$}}
      \put(1,77){\makebox(0,0)[c]{\scriptsize $3'$}}
      \put(1,105){\makebox(0,0)[c]{\scriptsize $2'$}}
      \put(36,126){\makebox(0,0)[c]{\scriptsize $1'$}}
      \put(63,126){\makebox(0,0)[c]{\scriptsize 0}}
      \put(90,126){\makebox(0,0)[c]{\scriptsize 1}}
    \end{picture}
    \caption{One of the possible vertically symmetric
      states of the FPL model for $n=7$}
    \label{gfplvs7}
  \end{minipage}
  \hspace{.05\linewidth}
  \begin{minipage}[t]{.425\linewidth}
    \centering
    \begin{picture}(97,127)
      \put(6,35){\scalebox{.8}{\includegraphics{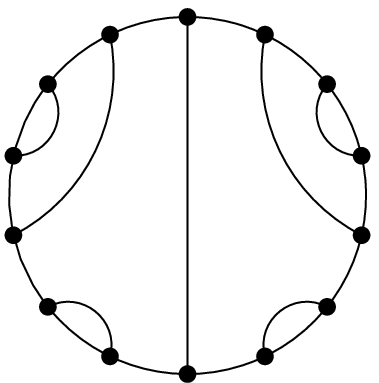}}}
      \put(49,128){\makebox(0,0)[c]{\scriptsize 0}}
      \put(70,124){\makebox(0,0)[c]{\scriptsize 1}}
      \put(87,110){\makebox(0,0)[c]{\scriptsize 2}}
      \put(97,90){\makebox(0,0)[c]{\scriptsize 3}}
      \put(97,68){\makebox(0,0)[c]{\scriptsize 4}}
      \put(87,48){\makebox(0,0)[c]{\scriptsize 5}}
      \put(70,35){\makebox(0,0)[c]{\scriptsize 6}}
      \put(50,30){\makebox(0,0)[c]{\scriptsize $0'$}}
      \put(28,35){\makebox(0,0)[c]{\scriptsize $6'$}}
      \put(10,48){\makebox(0,0)[c]{\scriptsize $5'$}}
      \put(1,68){\makebox(0,0)[c]{\scriptsize $4'$}}
      \put(0,90){\makebox(0,0)[c]{\scriptsize $3'$}}
      \put(10,110){\makebox(0,0)[c]{\scriptsize $2'$}}
      \put(28,124){\makebox(0,0)[c]{\scriptsize $1'$}}
      \put(6,5){\scalebox{.75}{\includegraphics{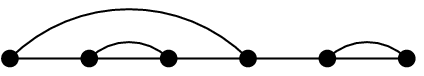}}}
      \put(8,0){\makebox(0,0)[c]{\scriptsize 1}}
      \put(25,0){\makebox(0,0)[c]{\scriptsize 2}}
      \put(43,0){\makebox(0,0)[c]{\scriptsize 3}}
      \put(60,0){\makebox(0,0)[c]{\scriptsize 4}}
      \put(77,0){\makebox(0,0)[c]{\scriptsize 5}}
      \put(94,0){\makebox(0,0)[c]{\scriptsize 6}}
    \end{picture}
    \caption{The pairing-pattern and the basis vector corresponding
      to figure \ref{gfplvs7}}
    \label{gpvs7}
  \end{minipage}
\end{figure}

Similarly to the previous case one sees that the bijection of the ASMs and
the states of the FPL model sends a vertically symmetric ASM into a
vertically symmetric state and vice versa, where we define a vertically
symmetric state of the FPL model in an evident way, see figure
\ref{gfplvs7} for an example.
Note that we use a special choice of numbered vertices and the numbering
which is more appropriate for the case under consideration.

The pairing-pattern corresponding to a vertically symmetric state of the
FPL model also has such a symmetry. We denote the set of such
pairing-patterns by $\Pi^{\mathrm V}_{2n+1}$, and the number of the
vertically symmetric ASMs corresponding to the pairing-pattern $\pi \in
\Pi^{\mathrm V}_{2n+1}$ by $A^{\mathrm V}_{2n+1}(\pi)$.

Now we establish a bijection of $\Pi^{\mathrm HT}_{2n+1}$ and the basis of
the dense O$(1)$ loop model identifying the $i$-th vertex with the $i'$-th
one and removing the pairing of the $0$-th vertex and the $0'$-th one, see
figure \ref{gpvs7}. Identifying the states of the dense O$(1)$ loop
model with the vertical symmetric pairing-patterns we formulate our second
new conjecture as follows.

\begin{con}
For the case of an even $N = 2n$ and the open boundary conditions the
vector
\[
\Psi = \sum_{\pi \in \Pi^{\mathrm{V}}_{2n+1}} \pi A^{\mathrm{V}}_{2n+1}
(\pi)
\]
is the ground state vector of the dense {\rm O(1)} loop model.
\end{con}

We have not succeeded to formulate a similar conjecture for the open case
with an odd~$N$.

Our results partially overlap the results of the recent paper by Pearce,
Rittenberg, and de~Gier \cite{PRG}. Namely, our conjecture 3 actually
coincides with the conjecture that the components of the ground state
vector for the open case and an even $N$ are given by the numbers of states
of the FPL model on the corresponding pyramid grid domain with specified
pairing-patterns.

{\it Acknowledgments}
The work was supported in part by the Russian Foundation for Basic Research
under grant \#01--01--00201 and the INTAS under grant \#00--00561.

\end{document}